 \documentclass[doublespacing]{elsart}

\usepackage{psfig}
\usepackage{amssymb}

\begin{document}

\begin{frontmatter}

\title{Baryon magnetic moments in  the background field method}

\author[GW]{F.X. Lee}
\author[GW]{R. Kelly}
\author[GW]{L. Zhou}
\author[BU]{W. Wilcox}

\address[GW]{Center for Nuclear Studies, 
Department of Physics,
The George Washington University, Washington, DC 20052, USA}
\address[BU]{Department of Physics, 
Baylor University, Waco, TX 76798, USA}

\begin{abstract}
We present a calculation of the magnetic moments for the baryon octet and decuplet
using the background-field method and standard Wilson gauge and fermion actions in the quenched approximation of lattice QCD.
Progressively smaller static magnetic fields are introduced on a $24^4$ lattice
at beta=6.0 and the pion mass is probed down to about 500 MeV.
Magnetic moments are extracted from the linear response of the masses to the background field.
\end{abstract}


\end{frontmatter}

 \section{Introduction}
%%%-------------------------------------------
Magnetic moments are an important fundamental property of particles. They
determine dynamical response of a bound system to a soft external stimulus, 
and provide valuable insight into internal strong interaction structure. 
Efforts to compute the magnetic moment on the lattice can be divided into two categories. 
One is from form factors which involves 
three-point functions~\cite{mart89,leinweber3,wilcox92,ji02,horse03,zanotti04,cloet04}. 
The other is the background field method using only two-point 
functions~\cite{mart82,bernard82,smit87,rubin95}.
The form factor method requires an extrapolation to $G_M(Q^2=0)$ that is potentially 
fallible due to the discrete momentum on the lattice~\cite{wilcox02}. 
The background field method, on the other hand, is free from this problem since 
a static magnetic field is applied to the gauge background.
Here we report a calculation in this method. It is an extension of our work 
on electric and magnetic polarizabilities~\cite{joe04,zhou02,zhou04}.

 \section{Method}
In order to place a magnetic field on the lattice, we construct an 
analogy to the continuum case. The fermion action is modified 
by the minimal coupling prescription 
\begin{equation}
D_\mu = \partial_\mu+gG_\mu + q A_\mu
\end{equation}
where $q$ is the charge of the fermion field and $A_\mu$ is the vector 
potential describing the background field. On the lattice, the prescription
amounts to a modified link variable 
\begin{equation}
U_\mu^\prime=U_\mu U_\mu^{(B)}.
\end{equation}
Choosing $A_y = B x $, a constant magnetic field B can be introduced 
in the $z$-direction.  Then the phase factor is in the y-links
\begin{equation}
U_y^{(B)}=\exp{(iqa^2Bx)}.
\end{equation}
On a finite lattice with periodic boundary conditions, to get a constant magnetic field, 
B has to be quantized by the condition
\begin{equation}
qBa^2={2\pi n \over N_x}, \hspace{1mm} n=1,2,3,\cdots
\end{equation}
to ensure that the magnetic phase factor is periodic in the x-direction.
However, for $N_x=N_y=16$ and $1/a=2 GeV$, the lowest field would give the proton 
a mass shift of about 390 MeV, which is too large (the proton is severely distorted). 
In this work, we abandon the quantization condition and choose to work with smaller fields.
To minimize the boundary effects, we work with fixed (or Dirichlet) b.c. in the x-direction and 
large $N_x$, so that quarks originating in the middle of the lattice have little chance of propagating to the edge.

We use $24^4$ lattice at $\beta=6.0$, and six kappa values $\kappa$=0.1515, 0.1525, 0.1535, 
0.1540, 0.1545, 0.1555, corresponding to pion mass of 1015, 908, 794, 732, 667, 522 MeV.
The critical value of kappa is $\kappa_c$=0.1571.
The strange quark mass is set at $\kappa$=0.1535. The source location for the quark propagators 
is (x,y,z,t)=(12,1,1,2).
We analyzed 87 configurations.
The following five dimensionless numbers 
$\eta=qBa^2$=+0.00036, -0.00072, +0.00144, -0.00288, +0.00576 give four small B fields 
(two positive, two negative) at 
$eBa^2$=-0.00108, +0.00216,  -0.00432, +0.00864 for both u and d (or s) quarks. 
Note that "small" here is in the sense that the mass shift is only a fraction of the proton mass:
$\mu B/m \sim$ 1 to 5\% at the smallest pion mass. 
In absolute terms, the field is enormous: $B\sim 10^{13}$ tesla.

The mass shift in the presence of small fields can be expanded as
a polynomial in B,
\begin{equation}  \label{mshift}
\delta m (B) = m(B)-m(0)= c_1 B + c_2 B^2+ c_3 B^3+c_4 B^4+\cdots
\end{equation}
To eliminate the contamination from the even-power terms, we calculate mass shifts 
both in the field $B$ and its reverse $-B$ for each value of $B$, 
then take the difference and divide by 2.
So in terms of cost, our calculation is equivalent to 11 separate spectrum calculations: 
5 original $\eta$ values, 5 reversed, plus the one at zero field to set the baseline.
In light of future dynamical simulations,
the factor can be reduced to 3 if only one nonzero field is desired.
Another benefit of repeating the calculation with the field reversed is that 
by taking the average of $\delta m (B)$ and $\delta m (-B)$ in the same dataset, 
one can eliminate the odd-powered terms in the mass shift. 
The coefficient of the leading quadratic term 
is directly related to the so-called magnetic dipole polarizability ($\beta$).

For a Dirac particle of spin $s$ in uniform fields, 
\begin{equation}
E_\pm=m_\pm\pm\mu B
\end{equation}
where the upper sign means spin up and the lower sign means spin-down, and 
$\mu=g {e\over 2m}s$. In our data, $m_{+}$ and $m_{-}$ correspond to the `11' and `22' 
diagonal components of the baryon correlation function in the absence of the background field. 
They are the same within statistical fluctuations. 
The average of the two components is usually used to extract the proton mass.
We tried the following three methods to 
extract the g factors, and found they are equivalent within statistical errors.
\begin{equation}
g=\left( {2m_+ m_- \over m_+ + m_-} \right){(E_+ - m_+)-(E_- - m_-)\over eBs}
\label{g1}
\end{equation}
\begin{equation}
g={m_+ (E_+ - m_+) - m_- (E_- - m_-)\over eBs}
\end{equation}
\begin{equation}
g={(E_+^2 - m_+^2) - (E_-^2 - m_-^2)\over 2eBs}
\end{equation}
The results quoted are from Eq.~(\ref{g1}).

 \section{Results and discussion}
\begin{figure}
\centerline{\psfig{file=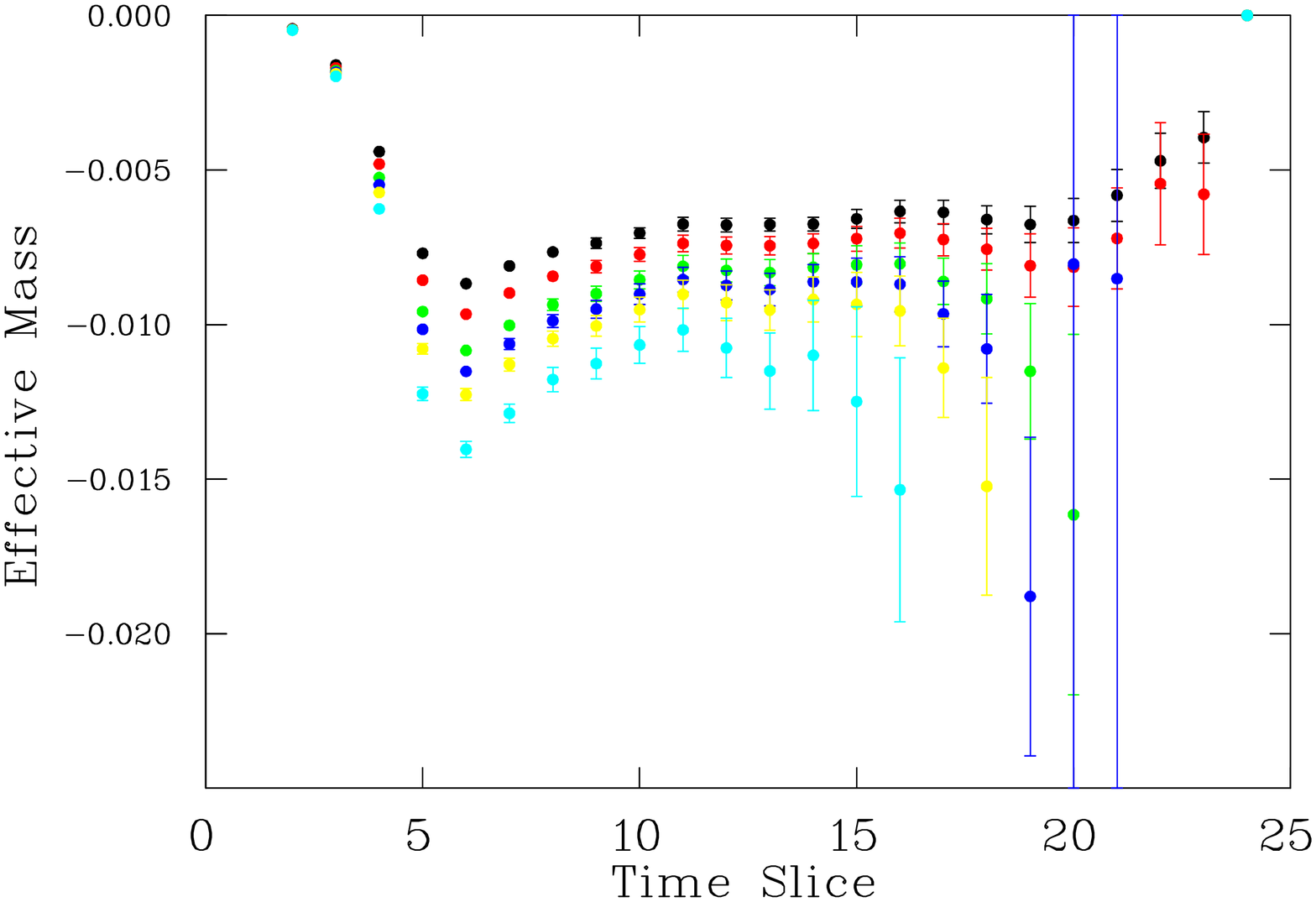,width=6.5in,angle=90}}
%\vspace*{-0.9cm}
\caption{Effective mass plot for proton mass shifts as the strongest magnetic field
in lattice units. Plotted is the combination $(E_+-m_+)-(E_--m_-)$ in Eq.~(\protect\ref{g1})
at all six quark masses in the order of decreasing quark mass from top to bottom.}
\label{emass-octp-shift-m4}
%\vspace*{-0.5cm}
\end{figure}

\begin{figure}
\centerline{\psfig{file=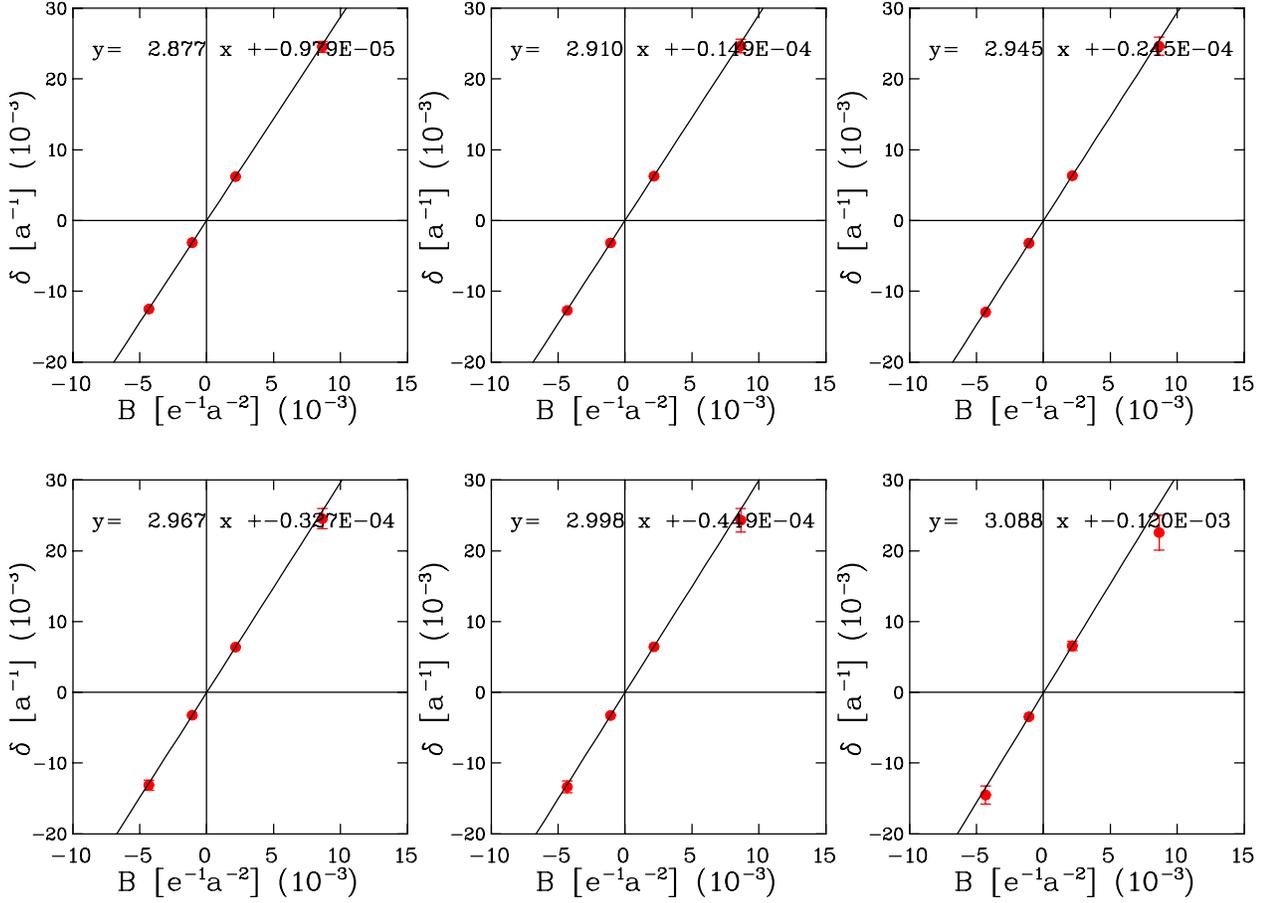,width=6.5in,angle=90}}
%\vspace*{-0.9cm}
\caption{Proton mass shifts as a function of magnetic field 
$B$ in lattice units at the six quark masses (lightest in the lower right corner). 
The slope of the mass shift at each quark mass gives the g factor corresponding to 
that quark mass.
The line is a fit using only the two smallest B values.}
\label{octp_linear_shift}
%\vspace*{-0.5cm}
\end{figure}

%%%-------------------------------------------
Fig.~\ref{emass-octp-shift-m4} displays a typical effective mass plot at the six kappa values for 
the strongest magnetic field for the proton. Good plateaus exist for all six quark masses.
Our results are extracted from the time window 12 to 15.
Fig.~\ref{octp_linear_shift} displays a typical mass shift, 
defined as $\delta=g (eBs)$ from Eq.~(\ref{g1}), as a function of the field
for the proton.
There is good linear behavior going through the origin for all the fields 
when the quark mass is heavy, 
an indication that contamination from the higher terms has been 
effectively eliminated. This is also confirmed numerically by the smallness of 
the coefficient in the $B^3$ term as shown in the same figure. 
At the lightest quark mass (lower right corner), there is a slight deviation from linear behavior.
For this reason, we only use the two smallest field values 
to do the linear fit at all the quark masses.

\begin{figure}
\centerline{\psfig{file=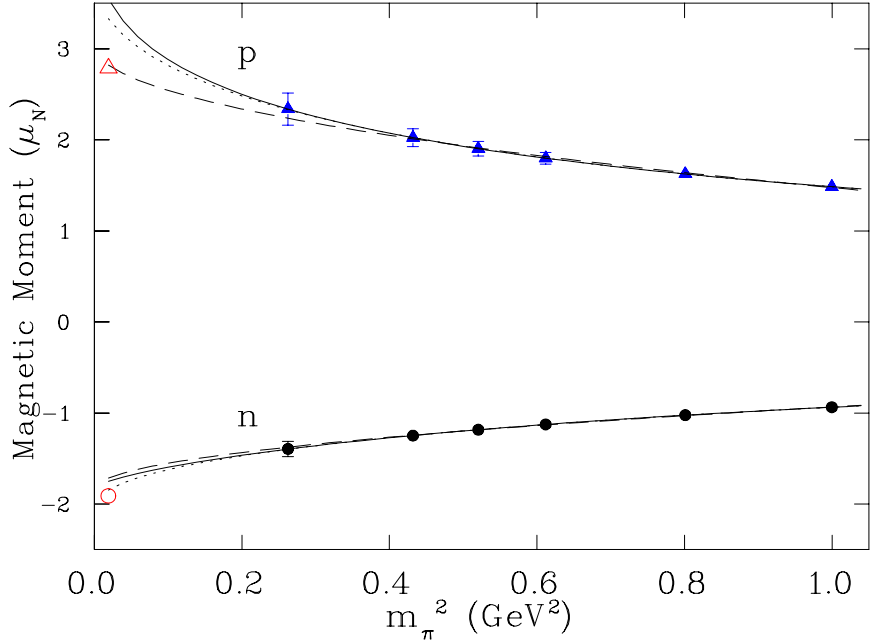,width=4.5in,angle=90}}
%\vspace*{-0.9cm}
\caption{Magnetic moments (in nuclear magnetons) for the proton and neutron 
as a function of $m_\pi^2$. The 3 lines are chiral fits according to 
Eq.~(\protect\ref{chiral1}) (dashed), Eq.~(\protect\ref{chiral2}) (dotted), and Eq.~(\protect\ref{chiral3}) (solid).
The experimental values, taken from the PDG~\protect\cite{pdg04}, 
are indicated by the empty symbols.}
\label{mag_pn}
%\vspace*{-0.5cm}
\end{figure}

Fig.~\ref{mag_pn} shows the results for the proton and the neutron as a function
of pion mass squared~\footnote{Since pion mass squared is proportional to
quark mass in QCD ($m_\pi^2 \varpropto m_q$), it is equivalent to plotting as a 
function of the quark mass.}.
Note that the g factors directly extracted from the data are in the particles natural 
magnetons.  To convert them to the commonly-used nuclear magnetons ($\mu_N$), we have scaled 
the results by the factor $938/M$ where $M$ is the mass of the particle
measured in the same calculation at each quark mass.
The line is a simple chiral fit using the ansatz
\begin{equation}
\mu=a_0 + a_1 m_\pi.
\label{chiral1}
\end{equation}
There is a lot of effort in the literature on magnetic
moments and their chiral extrapolations based on effective field 
theories~\cite{bora02,arndt02,savage02,beane03,beane04,hemm03,young04,leinweber04}.
Although different approaches produce the same chiral behavior 
near the chiral limit, there are issues surrounding the range of validity  
and model-independence away from the chiral limit.  
The lowest pion mass in our results (around 500 MeV) is probably too large for 
a meaningful application of the results from these studies.
Our goal here is to present the lattice data.
The simple ansatz in Eq.~(\ref{chiral1}) serves only to show that there is 
onset of non-analytic behavior as pion mass is lowered, so a linear extrapolation 
is probably not a good idea. 
To get some idea about the systematic uncertainty in the chiral fit, 
we tried two other different forms, one is 
\begin{equation}
\mu=a_0 + a_1 m_\pi+a_2 m_\pi^2
\label{chiral2}
\end{equation}
and the other the Pade form~\cite{leinweber04}
\begin{equation}
\mu=a_0/(1 + a_1 m_\pi+a_2 m_\pi^2).
\label{chiral3}
\end{equation}
The extrapolated value to the physical point is 3.04(6) for the proton and -1.84(3) 
for the neutron from Eq.~(\ref{chiral1}), 3.78(5) and -2.03(2) from Eq.~(\ref{chiral2}), 
and 4.21(5) and -1.84(2) from Eq.~(\ref{chiral3}).

\begin{figure}
\centerline{\psfig{file=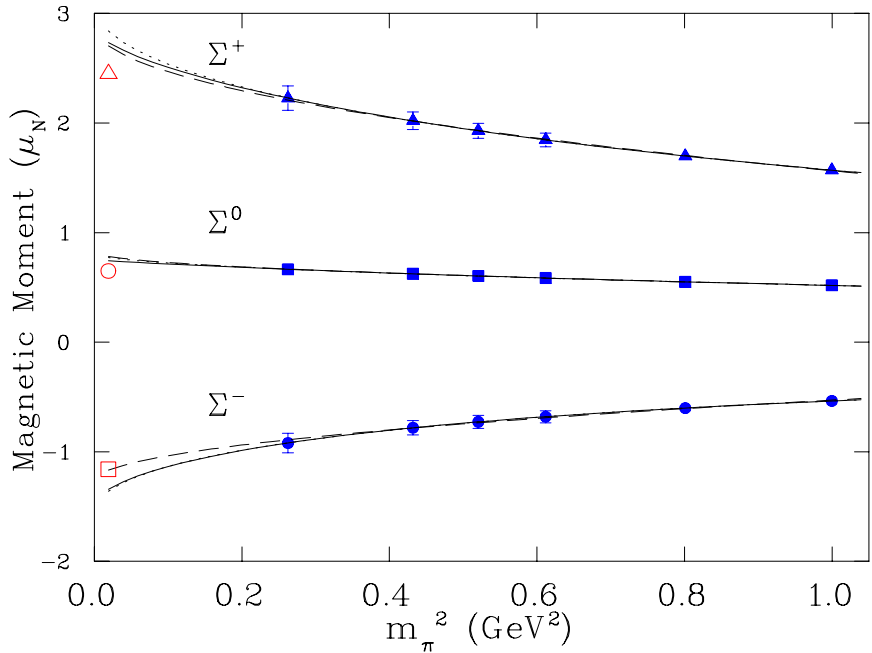,width=4.5in,angle=90}}
%\vspace*{-0.9cm}
\caption{Magnetic moments for the octet $\Sigma$.}
\label{mag_osig}
%\vspace*{-0.5cm}
\end{figure}

\begin{figure}
\centerline{\psfig{file=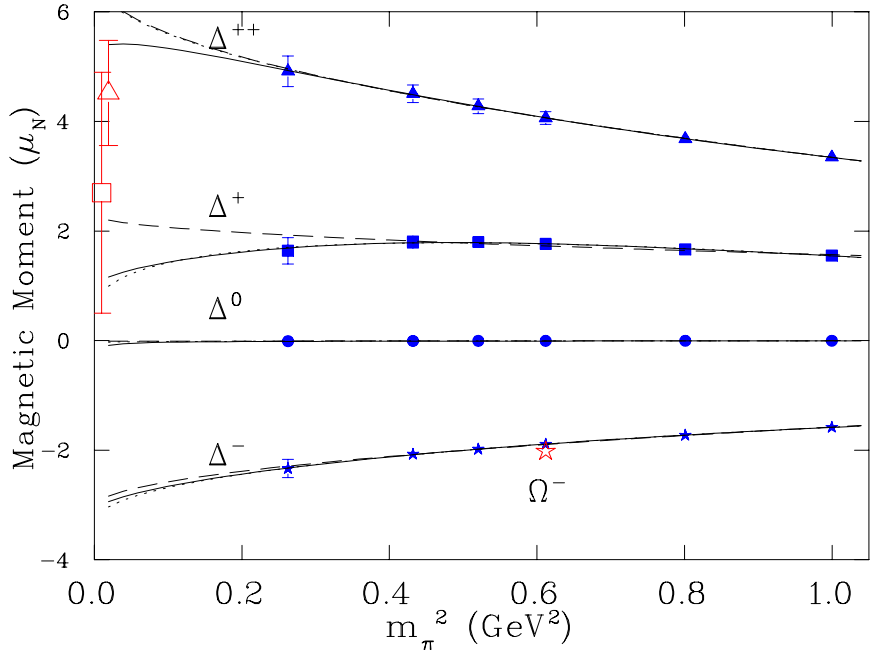,width=4.5in,angle=90}}
%\vspace*{-0.9cm}
\caption{Magnetic moments for the $\Delta$ states.}
\label{mag_decd}
%\vspace*{-0.3cm}
\end{figure}

\begin{figure}
\centerline{\psfig{file=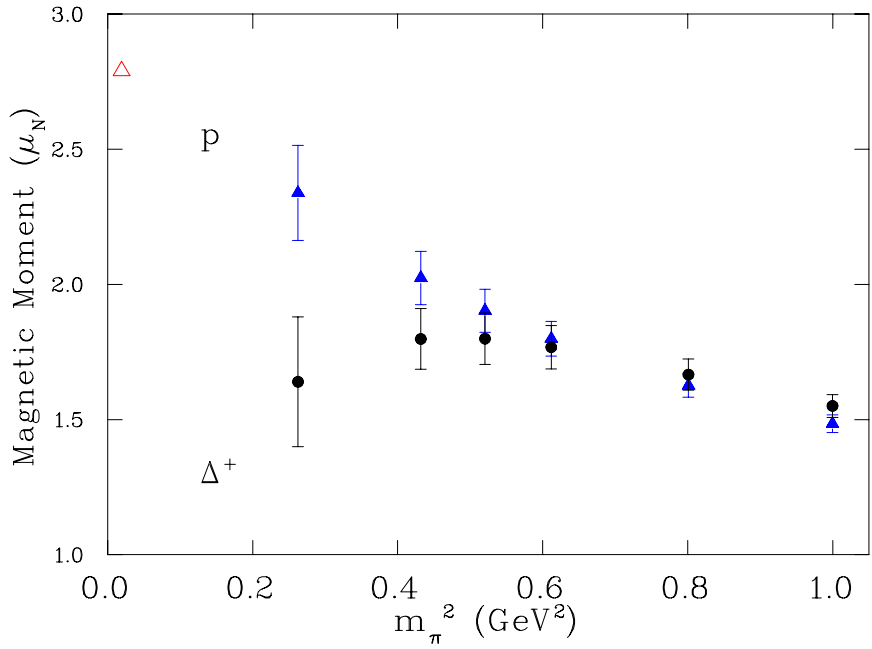,width=4.5in,angle=90}}
%\vspace*{-0.9cm}
\caption{Comparison of magnetic moments of the proton and $\Delta^+$.}
\label{mag_pdecdp}
%\vspace*{-0.3cm}
\end{figure}

Fig.~\ref{mag_osig} shows the results in the octet sigma channel.
The perfect agreement for the $\Sigma^-$ should be taken as a coincidence.
Fig.~\ref{mag_decd} shows the results for the four delta states.
The agreement with the well-known $\Omega^-$ moment in Fig.~\ref{mag_decd} 
is particularly encouraging since no adjustment is needed for this point 
(it is a direct comparison with the experimental value).
We take it as a sign of the correctness of our calculation.
The experimental numbers in the delta channel are not well-established.
The experimental value for $\Delta^+$ (slightly shifted for better view) is 
taken from Ref.~\cite{kotulla02}. 
For reader's convenience, all of our results are summarized in Table~\ref{mag_tab}.
Finally, Fig.~\ref{mag_pdecdp} shows the proton and $\Delta^+$ together.
The opposite curvatures are a signature of quenched chiral physics.
A similar behavior has been observed by ~\cite{zanotti03,young03} using 
the form factor method.

\begin{figure}
\centerline{\psfig{file=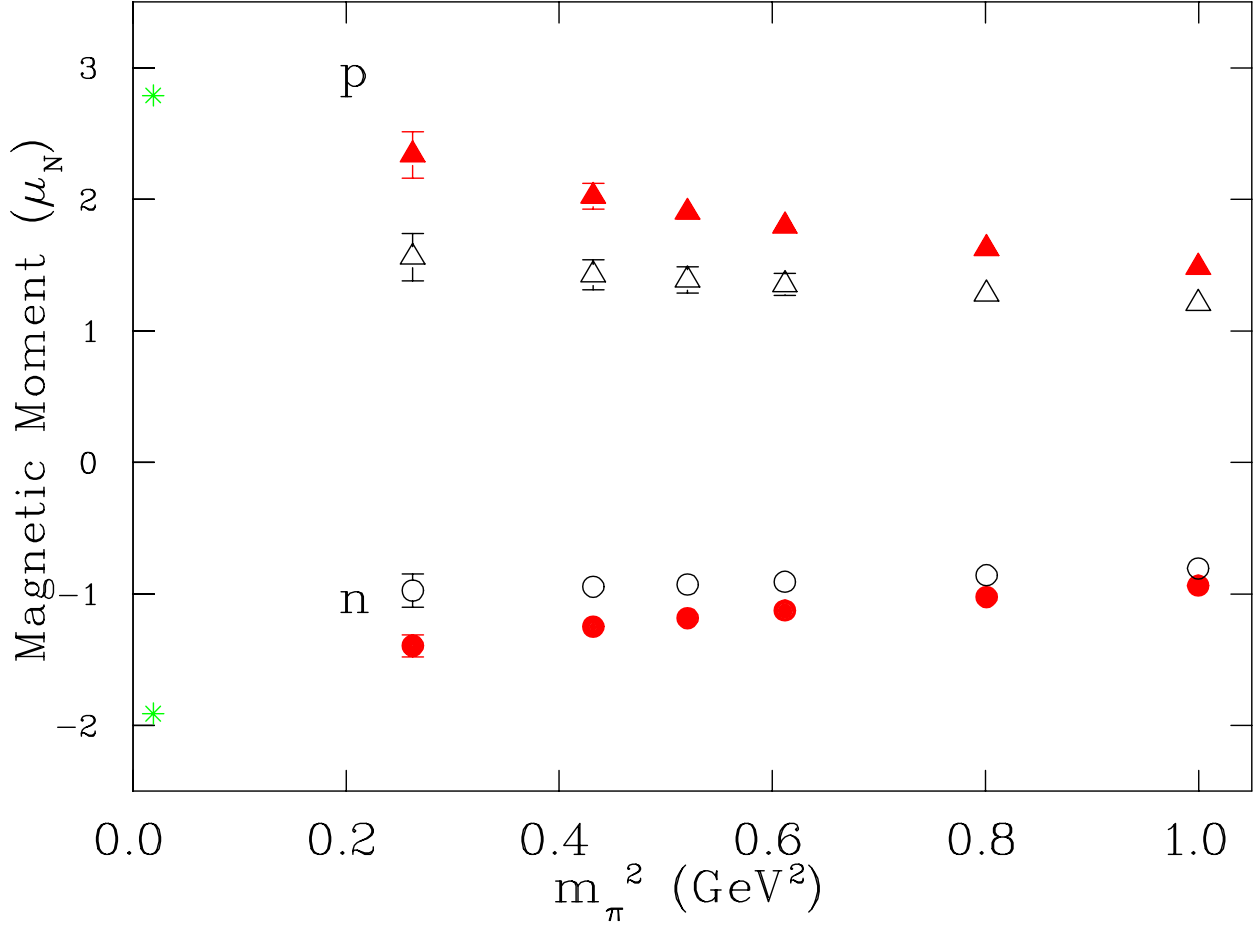,width=6.5in,angle=90}}
%\vspace*{-0.9cm}
\caption{Finite-volume effects on the magnetic moments of the proton and neutron.
The solid (red) symbols are from the $24^3\times 24$ lattice, and the empty symbols are
from the $16^3\times 24$ lattice. The experimental values are denoted by the stars.}
\label{mag_pn_2416_linear_12to15_12to15}
%\vspace*{-0.5cm}
\end{figure}

About the issue of finite-volume effects as related to the Dirichlet boundary conditions,
we did some further study. We repeated the entire calculation
on a smaller lattice $16^3\times 24$,
while keeping all the paraemeters the same as on the larger lattice
$24^3\times 24$. So the only difference is the spatial box size ($16^3$ vs. $24^3$).
Assuming the lattice spacing is the same (a=0.1 fm) on the two lattices, the
dimentionless combination $m_\pi L_x$ for the lightest pion is about 6.2 on the $24^3\times 24$,
and 4.2 on the $16^3\times 24$.
The effects on the magnetic moments of the proton and neutron are shown 
in Fig.~\ref{mag_pn_2416_linear_12to15_12to15}.
Relatively significant finite-volume effects are revealed, even at the heavier pion masses,
'spoiling' the trend towards the physical point.
We take this as indication that the $16^3\times 24$ lattice is too small.
To really pin down the finite-volume effects, we should repeat the calculation on a larger
lattice, say $32^3\times 32$. But it is currently beyond our computing resources.

 \section{Conclusion}
%%%-------------------------------------------
In conclusion, we have computed the magnetic moments of the baryon octet and 
decuplet on the lattice using the background field method and standard lattice 
technology. Overall, our results are consistent with experimental observations.
Detailed comparison with experiment must await a full account of systematic errors 
present in the results, such as finite-volume effects. 
In addition,  there is a need to push the calculations to smaller pion masses 
so that reliable chiral extrapolations can be applied.
Nonetheless, our results demonstrate that the method is robust and relatively cheap, 
as compared to the form factor method.  Only mass shifts are required. 
This may facilitate the push to smaller pion masses, perhaps with the help of chiral fermions 
(overlap, domain-wall, twisted mass, ...). 
Finally, we foresee no technical problems in doing dynamical (full-QCD) 
background-field calculations
in order to remove the effects of the quenched approximation.

%\section*{Acknowledgment}
%%%-------------------------------------------

This work is supported in part by U.S. Department of Energy
under grant DE-FG02-95ER40907, and by NSF grant 0070836.
The computing resources at NERSC and JLab have been used.

%
%%%%%%%%%%%%%%%%%%%%%%%%%%%%%%%%%%%%%%%%%%%%%%%%%%%%
%
%
%%%%%%%%%%%%%%%%%%%%%%%%%%%%%%%%%%%%%%%%%%%%%%%%%%%%
%
%
%%%%%%%%%%%%%%%%%%%%%%%%%%%%%%%%%%%%%%%%%%%%%%%%%%%%
%
%
%%%%%%%%%%%%%%%%%%%%%%%%%%%%%%%%%%%%%%%%%%%%%%%%%%%%%%%%%%%%%%%
%
%
%%%%%%%%%%%%%%%%%%%%%%%%%%%%%%%%%%%%%%%%%%%%%%%%%%%%%%%%%%%%%%%
%
%\begin{figure}
%\centerline{\psfig{file=mag_oxi_linear_13to16_13to16.ps,width=6.0in,angle=90}}
%\vspace*{-0.9cm}
%\caption{Magnetic moments for the octet $\Xi$.}
%\label{mag_oxi}
%\vspace*{-0.4cm}
%\end{figure}
%
%%%%%%%%%%%%%%%%%%%%%%%%%%%%%%%%%%%%%%%%%%%%%%%%%%%%%%%%%%%%%%%
%
%
%%%%%%%%%%%%%%%%%%%%%%%%%%%%%%%%%%%%%%%%%%%%%%%%%%%%%%%%%%%%%%%
%
%\begin{figure}
%\centerline{\psfig{file=mag_dsig_linear_12to15_12to15.ps,width=6.0cm,angle=90}}
%\vspace*{-0.9cm}
%\caption{Magnetic moments for the decuplet $\Sigma^*$.}
%\label{mag_dsig}
%\vspace*{-0.3cm}
%\end{figure}
%
%%%%%%%%%%%%%%%%%%%%%%%%%%%%%%%%%%%%%%%%%%%%%%%%%%%%%%%%%%%%%%%
%
%\begin{figure}
%\centerline{\psfig{file=mag_dxi_linear_12to15_12to15.ps,width=6.0cm,angle=90}}
%\vspace*{-0.9cm}
%\caption{Magnetic moments for the decuplet $\Xi^*$.}
%\label{mag_dxi}
%\vspace*{-0.3cm}
%\end{figure}
%
%%%%%%%%%%%%%%%%%%%%%%%%%%%%%%%%%%%%%%%%%%%%%%%%%%%%%%%%%%%%%%%
%
%
%%%%%%%%%%%%%%%%%%%%%%%%%%%%%%%%%%%%%%%%%%%%%%%%%%%%
%
%

%
%%%%%%%%%%%%%%%%%%%%%%%%%%%%%%%%%%%%%%%%%%%%%%%%%%%
\begin{center}
\begin{table}  % The * here means wide table across columns.
\caption{The computed magnetic moments for the baryon octet and decuplet 
in nuclear magnetons ($\mu_N$) as a function 
of the pion mass. The extrapolated values are based on Eq~(\protect\ref{chiral1}). 
The experimental values are taken from the PDG~\protect\cite{pdg04}.}
\label{mag_tab}
\begin{tabular}{llccccccc}
\hline
$\kappa$ & 0.1515 & 0.1525 & 0.1535 & 0.1540 & 0.1545 & 0.1555 & Extrap.   & Expt.  \\
$ m_\pi$ (GeV) & 1.015 & 0.908 & 0.794 & 0.732 & 0.667 & 0.522 &  &   \\
\hline
%\multicolumn{9}{l}{Baryon octet} \\
p            & 1.48(3) & 1.63(4) & 1.80(6) & 1.90(8) & 2.02(10) & 2.34(17) & 3.04(6) &  2.79  \\
n            &-0.94(2) &-1.02(2) &-1.13(3) &-1.18(4) &-1.25(5)  &-1.39(8)  &-1.84(3) & -1.91  \\
$\Sigma^+$   & 1.57(4) & 1.70(5) & 1.85(6) & 1.93(7) & 2.02(8)  & 2.23(11) & 2.87(3) &  2.45  \\
$\Sigma^0$   & 0.52(1) & 0.55(1) & 0.59(2) & 0.61(2) & 0.63(2)  & 0.67(3)  & 0.76(1) &  0.65  \\
$\Sigma^-$   &-0.54(4) &-0.60(4) &-0.68(6) &-0.73(6) &-0.78(7)  &-0.92(9)  &-1.48(5) & -1.16  \\
$\Xi^0$      &-1.07(4) &-1.10(4) &-1.13(4) &-1.15(4) &-1.17(5)  &-1.21(7)  &-1.37(1) & -1.25  \\
$\Xi^-$      &-0.76(6) &-0.77(7) &-0.77(8) &-0.77(8) &-0.77(9)  &-0.78(11) &-0.82(1) & -0.65  \\
$\Lambda$    &-0.59(2) &-0.60(2) &-0.60(2) &-0.61(2) &-0.61(2)  &-0.62(3)  &-0.70(1) & -0.61  \\
\hline
%\multicolumn{9}{l}{Baryon decuplet} \\
$\Delta^{++}$& 3.35(7) & 3.68(9) & 4.06(12)& 4.28(14)& 4.50(16) & 4.92(28) & 5.24(18)&  4.52(1.00)  \\
$\Delta^{+ }$& 1.55(4) & 1.67(6) & 1.77(8) & 1.80(10)& 1.80(11) & 1.64(24) & 0.97(8) &        \\
$\Delta^{0 }$&-0.002(0) &-0.003(0)&-0.004(1)&-0.005(1)&-0.007(1) &-0.011(6)&-0.035(2)&        \\
$\Delta^{- }$&-1.58(4) &-1.73(5) &-1.89(7) &-1.98(8) &-2.07(10) &-2.34(17) &-2.98(19)&        \\
$\Sigma*^+$  & 1.40(5) & 1.56(6) & 1.72(8) & 1.80(10)& 1.86(11) & 1.88(17) & 1.27(6) &        \\
$\Sigma*^0$  &-0.13(1) &-0.09(1) &-0.03(0) &-0.01(0) & 0.03(0)  & 0.08(2)  & 0.33(5) &        \\
$\Sigma*^-$  &-1.68(5) &-1.76(7) &-1.83(9) &-1.87(10)&-1.89(11) &-1.93(16) &-1.88(4) &        \\
$\Xi*^0$     &-0.09(2) &-0.06(1) &-0.022(5) &-0.002(1) & 0.018(6)  & 0.05(3)  & 0.16(4) &        \\
$\Xi*^-$     &-0.59(7) &-0.61(8) &-0.62(9) &-0.62(10)&-0.63(10) &-0.63(11) &-0.62(1) &        \\
\hline
\end{tabular}
\end{table}
\end{center}
%%%%%%%%%%%%%%%%%%%%%%%%%%%%%%%%%%%%%%%%%%%%%%%%%%%
%
\end{document}